\documentclass[twocolumn,prl,superscriptaddress]{revtex4}

\usepackage{graphicx}

\begin{document}

\title{Lower bound for electron spin entanglement from beamsplitter current correlations}

\author{Guido Burkard}
\address{IBM T.\ J.\ Watson Research Center, P.\ O.\ Box 218, Yorktown Heights, NY 10598}
\author{Daniel Loss}
\address{Department of Physics and Astronomy, University of Basel,
Klingelbergstrasse 82, CH-4056 Basel, Switzerland}

\newcommand{\1}{{{\bf S}_1}}
\newcommand{\2}{{{\bf S}_2}}

\newcommand{\p}{\mathbf{P}}

\newcommand{\Tr}{\mathrm{Tr}}
\newcommand{\T}{\mathrm{T}}

\newcommand{\bra}[1]{{\langle #1 |}}
\newcommand{\ket}[1]{{| #1 \rangle}}

\newcommand{\spup}{\ket{\!\uparrow}}
\newcommand{\spdown}{\ket{\!\downarrow}}
\newcommand{\spupup}{\ket{\!\uparrow\uparrow}}
\newcommand{\spupdown}{\ket{\!\uparrow\downarrow}}
\newcommand{\spdownup}{\ket{\!\downarrow\uparrow}}
\newcommand{\spdowndown}{\ket{\!\downarrow\downarrow}}

\newcommand{\bspup}{\bra{\uparrow\!}}
\newcommand{\bspdown}{\bra{\downarrow\!}}
\newcommand{\bspupup}{\bra{\uparrow\uparrow\!}}
\newcommand{\bspupdown}{\bra{\uparrow\downarrow\!}}
\newcommand{\bspdownup}{\bra{\downarrow\uparrow\!}}
\newcommand{\bspdowndown}{\bra{\downarrow\downarrow\!}}

\draft

\begin{abstract}
We determine a lower bound for the entanglement
of pairs of electron spins injected into a mesoscopic conductor.
The bound can be expressed in terms of experimentally accessible quantities,
the zero-frequency current correlators (shot noise power or
cross-correlators) after transmission through an electronic beam splitter.
The effect of spin relaxation ($T_1$ processes) and decoherence ($T_2$ processes)
during the ballistic coherent transmission of the carriers in the wires is taken
into account within Bloch theory.  The presence of a variable inhomogeneous
magnetic field allows the determination of a useful lower bound for
the entanglement of arbitrary entangled states.
The decrease in entanglement due to thermally mixed states is studied.
Both the entanglement
of the output of a source (entangler) and the relaxation ($T_1$)
and decoherence ($T_2$) times can be determined.
\end{abstract}

\maketitle

Quantum nonlocality has been an intriguing issue since the early days of quantum mechanics \cite{EPR}.
Nonlocal effects can come into play when a quantum system is composed of at least two subsystems
which are spatially separated.
Despite their simplicity, the Bell states of two distant quantum two-state subsystems (A and B)
\begin{eqnarray}
  \ket{\Psi_\pm} = \frac{1}{\sqrt{2}}\left(\spupdown \pm \spdownup\right),\label{eq:Psi}\\
  \ket{\Phi_\pm} = \frac{1}{\sqrt{2}}\left(\spupup \pm \spdowndown\right),\label{eq:Phi}
\end{eqnarray}
exhibit the essential phenomenology of quantum nonlocality 
(e.g., they violate Bell's inequalities \cite{Bell})
thus providing an ideal testing ground for quantum nonlocality.
Here, we represent the two-state systems as spins 1/2 with basis states
``spin up'' $\spup$ and ``spin down'' $\spdown$ with respect to an arbitrary fixed 
direction in space.

With the development of quantum information theory \cite{BD}, and in particular with
quantum communication, it has become clear that EPR pairs can also play the
role of a \textit{resource} for operations that are impossible with 
purely classical means.  In this context, two-state systems are referred
to as quantum bits (qubits), and quantum nonlocality is related to the
concept of entanglement (defined below).
A number of quantum information processes--quantum 
teleportation \cite{teleportation}, quantum key distribution \cite{qkd}, 
quantum dense coding \cite{qdc}, etc.--have been successfully
implemented using pairs of photons with entangled polarizations, 
i.e., in states such as Eqs.~(\ref{eq:Psi}) and (\ref{eq:Phi}).
Photons have the advantage that they are easily moved form
one place to another, allowing for experiments involving space-like
separations between detection events \cite{Bell}.
\begin{figure}[t]
\centerline{
\includegraphics[width=6cm]{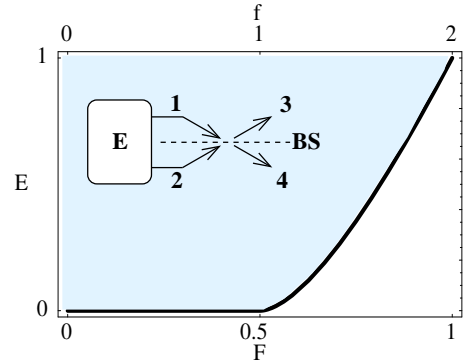}}
\caption{\label{fig:entangler}
Inset: Proposed setup with two-electron scattering at a beamsplitter (BS)
with transmittivity $T$.
Electrons are injected pairwise from the entangler (E) into contacts 1 and 2 
of the BS. The mean current $I=\langle I_\alpha\rangle$ 
and one of the current correlators $S_{\alpha\beta}$
are measured at the outgoing contacts $\alpha,\beta=3,4$.
Plot:  Entanglement of formation $E$ of the electron spins
versus singlet fidelity $F$ and the reduced correlator
$f=S_{33}/2eI T(1-T)$.
The curve illustrates the relation between noise and entanglement
for Werner states.  For general states, the curve represents a lower bound for the
entanglement, i.e.\ allowed values for $E$ and $f$ (or, equivalently, $F$) 
are represented by points in the shaded region.  Measuring
$f$ determines a lower bound for the entanglement $E$.}
\end{figure}

More recently, there has been increasing interest in the use of the spin of 
electrons in a solid-state environment for spin-based electronics \cite{Springer}
and as qubits for quantum computing \cite{LD}.
Subsequently, quantum communication on a mesoscopic scale, typically
on the order of micrometers in semiconductor structures (e.g.\ quantum wires), was
proposed \cite{MMM}.  Rather than achieving space-like
separation between detection events on the two sites (this would require
sub-picosecond detection), the idea here is to use quantum entanglement between 
parts of a coherently operating solid-state device (in the most extreme case,
a quantum computer).  It is then relevant to study the transport of
spin-entangled electrons in a many-electron system and possible means of
entanglement detection.  Two-particle interference
at a beam-splitter (BS) combined with the measurement of current fluctuations \cite{BLS}
(in general, the full counting statistics \cite{FCS}) was identified 
as a detector for entanglement.

In this paper, we go one step further, providing a \textit{lower bound} for 
the \textit{amount} $E$ of spin entanglement
carried by individual pairs of electrons, related to the zero-frequency current correlators
when measured in a BS setup (Fig.~\ref{fig:entangler}, Inset) 
by injecting the electrons separately
into the two ingoing leads (1 and 2) and measuring 
either the current autocorrelator $S_{\alpha\alpha}$
in one of the outgoing leads ($\alpha=3,4$) or the cross-correlator $S_{34}$.
It is assumed that the size of the
scattering region is smaller than both the coherence length and the mean
free path, allowing for ballistic and coherent transport.
In the following, $T$ will denote the transmittivity of the BS, 
i.e.\ the probability to be scattered from lead $1$ to lead $4$ (or from $2$ to $3$).
The ideal BS for the proposed setup does not give rise to backscattering (e.g.\ from
lead 1 back into lead 1, or from 1 into 2, etc.).  We will also analyze the effect of such 
backscattering processes, as they give rise to background shot noise
which is unrelated to entanglement.
During their transport, the electron spins will be exposed to decoherence
and relaxation due to spin-dependent scattering caused by magnetic impurities, nuclear
spins, or the spin-orbit coupling (see \cite{Flatte} for a review).
We include these effects within a Bloch equation formalism \cite{Slichter}.
Comparison between our theory and experiment will
(i) test proposed entanglers \cite{RSL,LMB,RL,Bena,Bouchiat,Oliver,SL}
and (ii) determine spin relaxation ($T_1$) and 
decoherence ($T_2$) times.

The materials and structures required for testing our theory, although at the
forefront of current capabilities, appear to be feasible.
The largest efforts seem to be necessary to realize
the electron spin entangler \cite{BLS} for which there exists a number of
theoretical ideas, using normal--\cite{RSL,LMB} or 
carbon-nanotube--superconductor junctions \cite{RL,Bena,Bouchiat},
or single \cite{Oliver}, or coupled quantum dots \cite{BLS,SL}.
The electronic BS and the measurement of BS current correlators
have been experimentally demonstrated in a GaAs/AlGaAs heterostructure \cite{Liu}.
Coherent transport of electron spins over more than $100\,\mu{\rm m}$ 
in GaAs has been observed \cite{Kikkawa}.

Traditionally, current correlations, and in particular the quantum partition (shot) noise
have been used to gain information about a 
scatterer beyond its conductance \cite{Blanter-Buttiker}.
Here, we use a known scatterer (the BS) to gain information
about the quantum state (more precisely, its entanglement) of the scattered particles.
The correlation
function between the currents $I_\alpha (t)$ and $I_\beta (t)$ in two leads
$\alpha,\beta = 1,..,4$ of the BS is defined as
\begin{equation}
  \label{eq:noise}
  S_{\alpha\beta}(\omega) = \lim_{\tau\rightarrow\infty}
  \frac{h\nu}{\tau}\int_0^\tau\!\!\!dt\,\,e^{i\omega t}
  \,{\rm Re}\,
  \Tr \left[ \delta I_\alpha(t)\delta I_\beta(0) \chi \right],
\end{equation}
where $\delta I_\alpha = I_\alpha - \langle I_\alpha\rangle$,
$\langle I_\alpha\rangle=\Tr ( I_\alpha \chi )$,
$\nu$ is the density of states in the leads, and $\chi$
is the density matrix of the injected electron pair
(below, we suppress the orbital part of $\chi$,
see \cite{BLS} for Coulomb effects).
Writing $\chi$ in the Bell basis, Eqs.~(\ref{eq:Psi}) and
(\ref{eq:Phi}), 
$\chi = F \ket{\Psi_-}\bra{\Psi_-} + G_0 \ket{\Psi_+}\bra{\Psi_+}
+ \sum_{i=\pm}G_i \ket{\Phi_i}\bra{\Phi_i}$,
and $S_{\alpha\beta}\equiv S_{\alpha\beta}(\omega = 0)$,
we arrive at
\begin{eqnarray}
  S_{\alpha\beta} &=&    F S_{\alpha\beta}^{\ket{\Psi_-}} + G_0 S_{\alpha\beta}^{\ket{\Psi_+}} 
                      + \sum_{i=\pm} G_i S_{\alpha\beta}^{\ket{\Phi_i}},
     \quad\quad \label{eq:noise-general-1}\\
  S_{\alpha\beta}^{\ket{\Psi}} & \equiv & \lim_{\tau\rightarrow\infty}
  \frac{h\nu}{\tau}\int_0^\tau\!\!\! dt \,{\rm Re}\,
  \bra{\Psi} \delta I_\alpha(t)\delta I_\beta(0) \ket{\Psi}.
     \label{eq:noise-general-2}
\end{eqnarray}
Using the standard scattering approach \cite{Blanter-Buttiker},
we have found earlier \cite{BLS} that the singlet state $\ket{\Psi_-}$ gives rise to enhanced
shot noise (and cross-correlators) at zero temperature,
$S^{\ket{\Psi_-}}_{33} = -S^{\ket{\Psi_-}}_{34}= 2eI T(1-T) f$,
with the \textit{reduced correlator} $f=2$, as compared to the
``classical'' Poissonian value $f=1$ \cite{noise-review}.
The average currents are given by $I=\langle I_3\rangle=\langle I_4\rangle = e/h\nu$.
We also know that all triplet states are noiseless,
$S^{\ket{\Psi_+}}_{\alpha\beta} = S^{\ket{\Phi_\pm}}_{\alpha\beta} = 0$ ($\alpha,\beta=3,4$).
Both the current autocorrelations (shot noise) and cross-correlations
are only due to the singlet component of the incident two-particle
wavefunction,
\begin{equation}
  \label{eq:Sf}
  S_{33}= -S_{34} = F S^{\ket{\Psi_-}} = 2eI T(1-T) f,
  \quad f=2F.
\end{equation}
Including backscattering with probability $R_B$, we find
\begin{eqnarray}
S_{33}  &=&  2eI  \left[\, 2 F (1 - R_B) \, T(1-T)  + R_B/2 \,\right], \\
S_{34}  &=&  - 2eI  \,    2 F (1 - R_B)  \, T(1-T) ,
\end{eqnarray}
where $I = (e/h\nu) (1-R_B)$.
Since $f^\prime = S_{34}/2eIT(1-T)=2F(1-R_B)\le f$
is smaller than $f$ without backscattering and 
the entanglement of formation $E$ is a monotonic function of $f$ (see below and 
Fig.~\ref{fig:entangler}), we still obtain a lower bound on $E$
(the bound will become less informative as $R_B$ increases).
Note that this does not hold for the autocorrelator $S_{33}$.
However, one can determine $R_B$, e.g.\ by measuring the shot noise power using
normal Fermi lead inputs \cite{Liu} and then obtain $f$ from either $S_{33}$ or $S_{34}$.

The entanglement of a bipartite state $\chi\in{\cal H}_A\otimes{\cal H}_B$
can be quantified by its entanglement of formation \cite{BDSW}
$E(\chi)=\min_{\{(\ket{\chi_i},p_i)\}\in \cal{E}(\chi)} \sum_i p_i S_N(\ket{\chi_i})$,
where $\cal{E}(\chi)$ denotes the set of ensembles $\{(\ket{\chi_i},p_i)\}$ 
for which $\chi=\sum_i p_i\ket{\chi_i}\bra{\chi_i}$.
We have used the von Neumann entropy of the reduced density matrix
$\rho_B=\Tr_A\ket{\psi}\bra{\psi}$,
$S_N(\ket{\psi}) = -\Tr_B \rho_B \log\rho_B$
(logarithms are in base 2).
A state with $E>0$ ($E=1$) is (maximally) entangled, whereas a state
with $E=0$ is separable (in the case of a pure state, it is a product $\psi_A\otimes\psi_B$).
The Bell states Eq.~(\ref{eq:Psi}) and (\ref{eq:Phi}) are maximally entangled.
Neither local operations nor classical communication (LOCC) between subsystems A and B
can increase $E$.  In quantum information
theory, $E(\chi)$ is the maximal ratio $N/M$ of the number $N$ of EPR 
pairs (maximally entangled states) required to form $M$ copies of $\chi$
as $N\rightarrow\infty$;
$E$ is the quantity that measures \textit{how much} of the
resource (quantum entanglement) is available.

\begin{figure}[t]
\centerline{
\includegraphics[width=8.5cm]{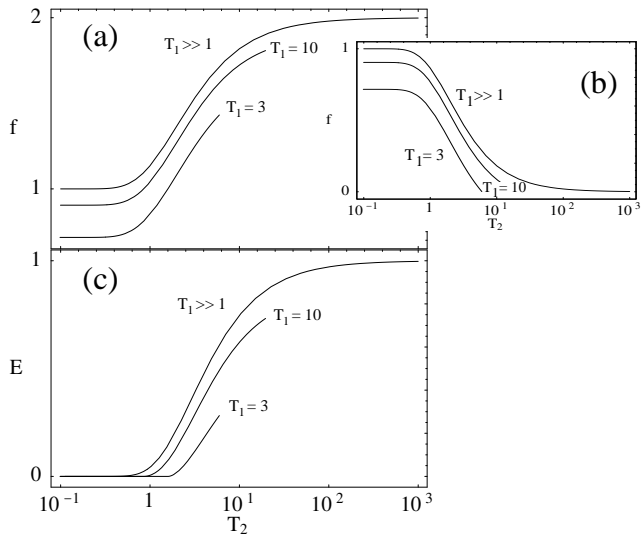}}
\caption{\label{fig:plots}
Homogeneous magnetic field $\delta h=0$ and $\tilde{P}=1$.
(a) $f$ of the spin singlet state $\ket{\Psi_-}$ after
ballistic transmission through a BS as a function of
the spin decoherence time $T_2$, in units of the ballistic transmission
time $t_0=L/v_F$.  Different curves correspond to different values
of the spin relaxation time $T_1$ (same units).  Note that we
only plotted the curves for $T_2\le 2T_1$.
(b) $f$ for a spin triplet state $\ket{\Psi_+}$.
Since $f\le 1$, the lower bound on entanglement is zero, i.e.\
we cannot learn anything about entanglement of injected triplets
at $\delta h=0$.
(c) Lower bound on the entanglement of formation $E$.}
\end{figure}
For arbitrary $\chi$, $E(\chi)$ cannot be expressed as a function
of only its singlet fidelity $F=\bra{\Psi_-}\chi\ket{\Psi_-}$.
However, this is possible for the so-called Werner states \cite{Werner}
\begin{equation}
  \label{werner}
  \rho_F \!=\! F \ket{\Psi_-}\bra{\Psi_-}  \!+\!  \frac{1\! -\! F}{3}
\!\left( \ket{\Psi_+}\bra{\Psi_+}\!+\! \sum_{i=\pm}\ket{\Phi_i}\bra{\Phi_i}\right)\! ,
\end{equation}
being the unique rotationally invariant states with singlet fidelity $F$.  
It is known \cite{BDSW} that
$E(F)\equiv E(\rho_F)=H_2(1/2 +\sqrt{F(1-F)})$ 
if $1/2<F\le 1$ and $E(F)\equiv E(\rho_F)=0$ if $0 \le F < 1/2$,
with the dyadic Shannon entropy $H_2(x)=-x\log x -(1-x)\log(1-x)$.
Together with Eq.~(\ref{eq:Sf}), this enable us to express
the entanglement of $\rho_F$ in terms
of the reduced correlator $f$ (Fig.~\ref{fig:entangler}).

We generalize this result to arbitrary mixed states $\chi$ of 
two spins (qubits).
Any state $\chi$ can be transformed into the Werner state $\rho_F$ with the same
singlet fidelity $F$ by a random bipartite rotation \cite{BBPSSW,BDSW},
i.e.\ by choosing a random $U\in SU(2)$ and applying $U\otimes U$ to $\chi$.  Since this 
operation involves only LOCC, $E$ cannot increase,
\begin{equation}
  \label{eq:bound}
  E(F)\le E(\chi).
\end{equation}
The entanglement of formation $E(F)$ of the corresponding Werner state therefore provides 
a \textit{lower bound} on the entanglement of $\chi$ (Fig.\ref{fig:entangler}).
Thus, a noise signal exceeding
the Poissonian limit ($f>1$) in the BS setup can in principle be interpreted as a
sign of entanglement between the electron spins injected into leads 1 and 2
\footnote{If the current is carried by quasiparticles
with charge $e^*$, $f$ will be renormalized by a factor $e/e^*$.}.

We now include relaxation and decoherence into our analysis.
At time $t=0$, we start with a spin singlet (upper sign) or triplet 
(lower sign) state
\begin{equation}
  \label{eq:dmatrix}
  \chi (0) = \ket{\Psi_\mp}\bra{\Psi_\mp}.
\end{equation}
We describe the dynamics of $\chi(t)$ in a
field $\mathbf{B}\parallel \hat{z}$ and in the presence of
spin decoherence ($T_2$ processes) and relaxation ($T_1$) 
phenomenologically within a single-spin Bloch equation for the polarization
$\p = (\langle \sigma_x\rangle,\langle \sigma_y\rangle,\langle \sigma_z\rangle )$,
\begin{equation}
  \label{eq:Bloch}
  \dot{\p} = \p \times \mathbf{h} - R (\p-\tilde{\p}) \equiv  -\Omega (\p - \tilde{\p}),
\end{equation}
with $\langle\sigma_i\rangle = \Tr(\sigma_i \rho)$,
$\mathbf{h} = g\mu_B \mathbf{B} = (0,0,h)$, 
the stationary polarization $\tilde{\p} = (0,0,\tilde{P})$
(note that $\tilde{\p}\times\mathbf{h}=0$), and the relaxation matrix
\footnote{Within standard weak-coupling theory, 
$T_2 = [(2T_1)^{-1} + T_\parallel^{-1}]^{-1}\le 2 T_1$
($T_\parallel^{-1}$ is the longitudinal decoherence rate).}
$R_{ij}=\delta_{ij}R_i$ with $R_1=R_2=T_2^{-1}$ and $R_3=T_1^{-1}$.
Solving Eq.~(\ref{eq:Bloch}), we obtain
\begin{equation}
  \label{eq:Bsolution}
  \p(t) = e^{-\Omega t}\p (0)+(1-e^{-\Omega t})\tilde{\p},
\end{equation}
or, in terms of the spin density matrix,
\begin{equation}
  \label{eq:superop0}
  \rho (t) = (P_0+\p (t)\cdot\mbox{\boldmath $\sigma$})/2 \equiv \Lambda_h (t) [\rho (0)],
\end{equation}
with the superoperator ($a(t) = 1- e^{-t/T_1}$)
\footnote{The superoperator $\Lambda_h (t)$ is
linear and trace-preserving on density matrices with arbitrary trace
since we have not imposed the trace condition $P_0=1$ at this point.}
\begin{widetext}
\begin{equation}
  \label{eq:superop1}
  \Lambda_h (t) [\rho] = \left(\begin{array}{c c}
\frac{1}{2}(\rho_{\uparrow\uparrow}+\rho_{\downarrow\downarrow})(1+a(t)\tilde{P})
+\frac{1}{2}(\rho_{\uparrow\uparrow}-\rho_{\downarrow\downarrow})e^{-t/T_1} &
e^{-t/T_2 + iht}\rho_{\uparrow\downarrow} \\
e^{-t/T_2 - iht}\rho_{\downarrow\uparrow} &
\frac{1}{2}(\rho_{\uparrow\uparrow}+\rho_{\downarrow\downarrow})(1-a(t)\tilde{P})
-\frac{1}{2}(\rho_{\uparrow\uparrow}-\rho_{\downarrow\downarrow})e^{-t/T_1}
\end{array}\right),
\end{equation}
\end{widetext}
with the matrix elements $\rho_{ij}=\langle i|\rho|j \rangle$
and $(i,j=\uparrow,\downarrow)$.
We apply $\Lambda_h (t)$ to both spins individually,
\begin{equation}
  \label{eq:superop2}
  \chi(t) = \left(\Lambda_{h_1}(t)\otimes\Lambda_{h_2}(t)\right) [\chi(0)],
\end{equation}
where $h_i$ is the field at electron $i$.
Using Eq.~(\ref{eq:Sf}) and $F(t) = \bra{\Psi_-}\chi(t)\ket{\Psi_-}$ at
time $t\ge 0$, we obtain
\begin{eqnarray}
  \label{result}
  f(t) &=& \pm e^{-2t/T_2} \cos(\delta h \, t)\nonumber\\
              &&   + \frac{1}{2}(1+e^{-2t/T_1}) 
                   - \frac{1}{2} (1-e^{-t/T_1})^2 \tilde{P}^2,
\end{eqnarray}
where $\delta h=h_1-h_2$.
If the decoherence time $T_2^{(1,2)}$ of the two electrons is
different, then $T_2$ in Eq.~(\ref{result}) becomes
$T_2^{\rm EPR}=(1/T_2^{(1)}+1/T_2^{(2)})^{-1}$.  
We define $T_1^{\rm EPR}$ similarly if $\tilde{P}=0$.
However, if $\tilde{P}=1$, then
$\exp(-t/T_1)$ is replaced by $\exp(-t/T_1^{(1)})+\exp(-t/T_1^{(2)})$.

A homogeneous field, $\delta h=0$, does not affect $f$.
For slow relaxation, $T_1\gg t$, we find $f(t)= 1\pm e^{-2t/T_2}$.
In Fig.~\ref{fig:plots}a we plot $f$ for $\delta h=0$ and 
$\tilde{P}=1$ versus $T_2$ in units of the ballistic transmission
time \cite{noise-review} $t_0=L/v_F$ ($L$=length of ballistic trajectory, $v_F$=Fermi
velocity).
For unentangled triplet states, 
$\chi (0) = \spupup\bspupup,\spdowndown\bspdowndown$,
we find $f\le 1/2$ for
all $T_1$, $T_2$, and $\tilde{P}$ (Fig.~\ref{fig:plots}b).

\begin{figure}
\centerline{
\includegraphics[width=8.8cm]{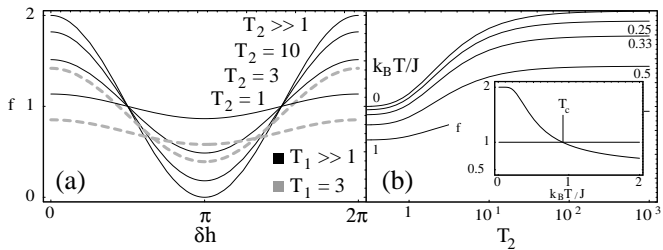}}
\caption{\label{fig:field}
(a) Reduced current correlator $f$ versus field inhomogeneity $\delta h=h_1-h_2$ 
(in units of $\hbar/t_0 g\mu_B$, $t_0$=ballistic transmission time,
$g$=g-factor, $\mu_B$=Bohr magneton) for an injected singlet state
and $\tilde{P}=1$.
Solid lines represent $T_1\gg 1$ and $T_2=1,3,10,\infty$, grey
dashed lines $T_1=3$ and $T_2=1,3$.
For injected triplets, the plot is phase-shifted
by $\pi$, providing tight lower bounds at $\delta h=\pi$.
Tight lower bounds for \textit{any} input state can
be determined by varying the direction of $\delta{\mathbf h}$.
(b) Plot of $f$ for a thermally mixed initial state
versus $T_2$ (in units of $t_0$)
for $T_1\gg t$ and $\tilde{P}=1$.
The various curves correspond to
$k_B \T/J = 0, 0.25, 0.33, 0.5, 1$, where $J$ and $\T$ denote the
exchange energy and temperature during the preparation of the state.
Inset:  The maximal $f$ (at $T_1, T_2\gg t_0$) versus $k_B \T/J$.  
There is no entanglement ($f\le 1$, $E=0$) above the critical 
temperature $\T_c = 0.91\,J/k_B$.}
\end{figure}

An inhomogeneous field $\delta h\neq 0$ (or, equivalently, a local
controllable Rashba spin-orbit coupling \cite{EBL}) has the effect
of continuously rotating singlets into triplets and vice versa
(Fig.~\ref{fig:field}).  This a lower bound of the triplet 
entanglement, $\chi(0)=\ket{\Psi_+}\bra{\Psi_+}$,
which is as tight as Eq.~(\ref{eq:bound}) for the singlet,
\begin{equation}
  \label{eq:bound-general}
  E \ge \max_{\delta {\bf h}}  E\left(f\left(\delta {\bf h}\right)/2\right),
\end{equation}
where $f(\delta {\bf h})$ is the measured noise power (or cross-correlator),
$E(F)$ is the entanglement of the Werner state 
$\rho_F$ (Fig.~\ref{fig:entangler}).  If a field
inhomogeneity $\delta {\bf h}$ can be created pointing in \textit{arbitrary} directions
in space, then the above result represents a tight lower bound
for \textit{any} injected entangled state.  In particular, each
maximally entangled state $\ket{\Psi}$ will be detected in this way, since
there exists a $u=\exp(-i\delta {\bf h}\cdot\mbox{\boldmath $\sigma$})\in SU(2)$
such that $u^\dagger\otimes u \ket{\Psi} = \ket{\Psi_-}$.
This rotation can also be done unilaterally, i.e.\ there is
a $v\in SU(2)$ with $v\otimes 1\ket{\Psi} = \ket{\Psi_-}$
(see also \cite{EBL}).

Finally, we study the case where the spin state of the injected
pair of carriers Eq.~(\ref{eq:dmatrix}) is mixed, because it
is prepared at a temperature $\T$ comparable to the energy splitting
between spin states, typically (if the Zeeman effect is negligible)
the exchange energy $J$, i.e.\ the singlet-triplet splitting.
In this case, $\chi(0)=\rho_F$ with 
$F=(1+3e^{-J/k_B \T})^{-1}$ where $k_B$ is Boltzmann's constant.
We only show the resulting $f$ for $T_1\gg t$ here (the full expression will be reported
elsewhere \cite{unpublished}),
\begin{equation}
  \label{eq:norelax}
  f(t)=\frac{1+e^{-2t/T_2}
          + e^{-J/k_B \T}\left( 1-e^{-2t/T_2}\right) }{1+3e^{-J/k_B \T}},
\end{equation}
which is the statistical mixture of Eq.~(\ref{result}) for
the singlet and triplet with the
appropriate Boltzmann weights (Fig.~\ref{fig:field}b).
Above the critical temperature $\T_c = 0.91\,J/k_B$ there is no
entanglement even for $T_1, T_2\rightarrow \infty$.

\textit{Acknowledgments:}  We thank D. P. DiVincenzo
and B. M. Terhal for valuable discussions.
DL acknowledges funding from the Swiss NSF,
NCCR Nanoscience, and DARPA QUIST and SPINS.

\end{document}